\documentclass[letterpaper]{article} 
\usepackage[preprint]{aaai2027}  
\usepackage[hyphens]{url}  
\usepackage{graphicx} 
\usepackage{natbib}  
\usepackage{caption} 
\usepackage{algorithm}
\usepackage{algorithmic}

\usepackage{newfloat}
\usepackage{listings}
\DeclareCaptionStyle{ruled}{labelfont=normalfont,labelsep=colon,strut=off} 
\floatstyle{ruled}
\newfloat{listing}{tb}{lst}{}
\floatname{listing}{Listing}

\usepackage{booktabs}

\usepackage{makecell}

\usepackage{amsmath}
\usepackage{multirow}
\usepackage{subcaption}

\usepackage{makecell}

\usepackage{amsmath}
\usepackage{multirow}
\title{CogenPVG: \textbf{Cog}nitive-\textbf{En}hanced Reflective Multi-Agent Framework for \textbf{P}ersuasive \textbf{V}ideo \textbf{G}eneration}
\author{
    Yuntian Xiao\textsuperscript{\rm 1,\rm 2}, Shoulong Zhang\textsuperscript{\rm 2}, Wenfeng Song\textsuperscript{\rm 3},\\
    Yan Wang\textsuperscript{\rm 2}, Yi Chen\textsuperscript{\rm 4}, Shuai Li\textsuperscript{\rm 1}\corresponding
}
\affiliations{

    \textsuperscript{\rm 1}Beihang University, \textsuperscript{\rm 2}Zhongguancun Laboratory,\\ \textsuperscript{\rm 3}Beijing Information Science and Technology University, \\
    \textsuperscript{\rm 4}Beijing Technology and Business University

}

\begin{document}

\maketitle

\begin{figure*}[t]
    \centering
    \includegraphics[width=0.9\linewidth]{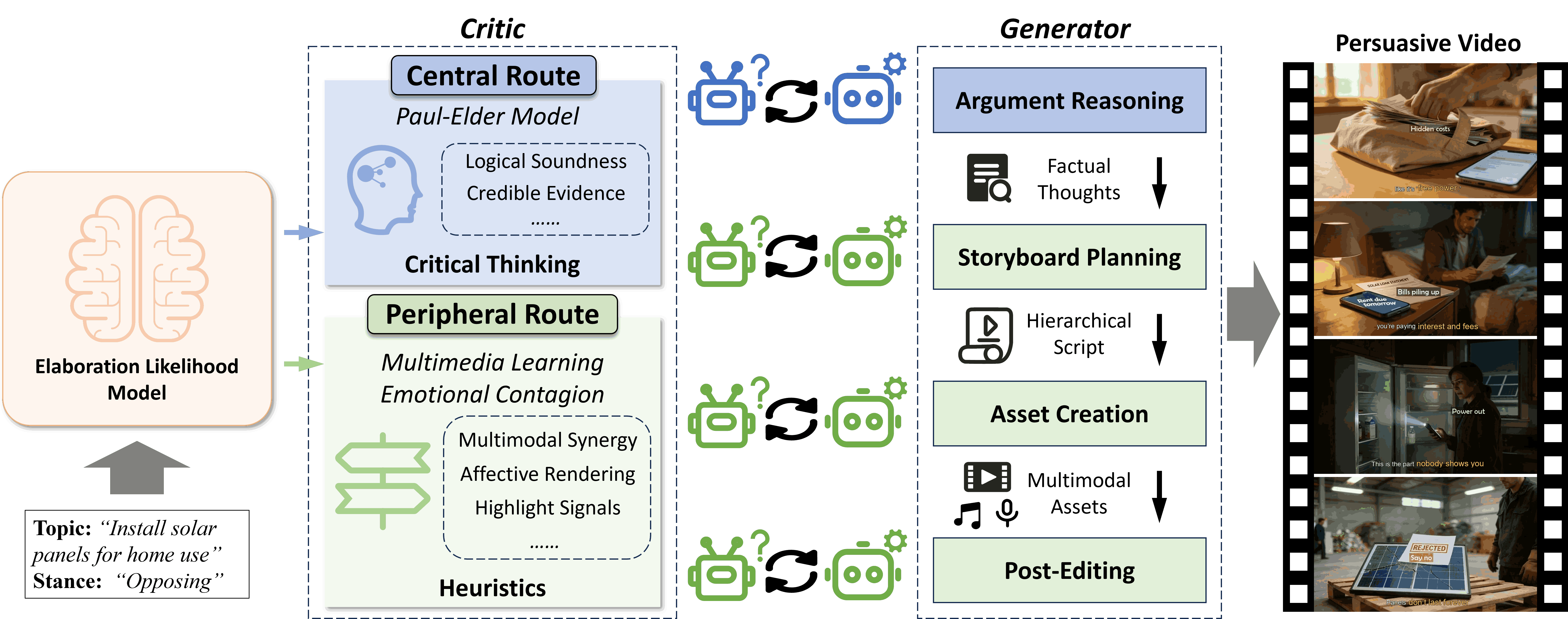}
    \caption{Overview of our CogenPVG. We incorporate the well-established dual-route persuading model of ELM for persuasive video generation.}
    \label{fig:teaser}
\end{figure*}

\begin{abstract}
  Persuasive video generation (PVG) is a valuable yet under-explored research topic. Despite the significant advances in multimodal content generation, AI-empowered automated creation of human-made-like videos with substantial persuasiveness remains a formidable challenge. In this paper, we propose CogenPVG, a novel \textbf{Cog}nitive-\textbf{en}hanced reflective multi-agent framework tailored for \textbf{P}ersuasive \textbf{V}ideo \textbf{G}eneration task.
  Given the topic and stance from the user, we decouple the sophisticated generation process into four sequential stages: argument reasoning, storyboard planning, asset creation, and post-editing, imitating the workflow of human video producers. To ensure high persuasiveness, each stage is equipped with a pair of generator and critic agents, following a reflective refinement scheme grounded in a solid psychological theory of persuasion, the Elaboration Likelihood Model (ELM). In the argument reasoning stage, we generate highly logical and credible reasoning thoughts under the guidance of critical thinking theory, enabling cognitive enhancement via the central route of the ELM. For the other three stages, we generate and optimize multimodal assets, assembling them into a persuasive video guided by theories of heuristics, as the peripheral route of the ELM.
  To the best of our knowledge, CogenPVG is the first work focused on general persuasive topics, without being confined to commercial purposes. Extensive experiments and comprehensive analysis demonstrate that our framework achieves the best persuasion performance, thereby proving the effectiveness of our proposed multi-agent framework for the PVG task.
\end{abstract}



\section{Introduction}
Persuasive videos circulate widely across social networks and media platforms (e.g., TikTok, YouTube, Douyin), wielding considerable influence due to their purpose of changing or reinforcing individual attitudes and beliefs~\cite{political, political2, advocacy}. As an enriched information medium, persuasive video creation has garnered the interest of researchers in commercial advertising~\cite{ad, ad2} and holds considerable potential for education~\cite{edu}, news commentary~\cite{news}, and debunking misinformation~\cite{debunk}. However, persuasive video generation (PVG) is a challenging and interdisciplinary task that integrates generative AI technologies with cognitive science and psychological theories, yet it remains an under-explored area. Considering its significance for both academic research and daily life, we propose the first practical solution toward AI-empowered persuasive video generation.

Prior video generation methods leverage large language models (LLMs)~\cite{gpt, gemini, deepseek} to build agents for dedicated multi-step planning,
followed by generating multimodal assets and assembling into a long-form video~\cite{videoauteur, vlogger, animaker, anim_director, mmstory, movieagent, aesopagent}, taking advantage of powerful generative models for a range of modalities such as image~\cite{storydiffusion, sdxl, seedream}, video clip~\cite{wan, cogvideox, seedance} and human speech~\cite{cosyvoice, seed_tts}. Although existing studies can produce high-quality videos and provide valid multi-agent frameworks, they are mainly confined to narrative-centered video types, such as story visualization~\cite{mmstory, animaker, anim_director}, movies~\cite {movieagent}, and vlogs~\cite{personavlog, vlogger}, whose primary concern lies in the consistency of characters, richness of scenes, and vividness of plots, while ignoring their cognitive impact, such as persuasiveness.

\begin{table*}[t]
    \centering
    \tiny
    {\fontsize{8}{\baselineskip}\selectfont
\setlength{\tabcolsep}{1mm}
    \begin{tabular}{@{}ccccc@{}}
        \toprule
ELM Route &  Psychological Theory & Applied Model & CogenPVG Stage & Criteria \\
        \midrule
Central  & Critical Thinking & Paul-Elder & AR & \makecell[l]{relevance, depth, breath, significance, accuracy, precision, \\ sufficiency, clarity, logic}  \\
        \midrule
\multirow{3}{*}[-1.5ex]{Peripheral} & \multirow{3}{*}[-1.5ex]{\makecell[c]{Fluency Heuristic \\Affect Heuristic}} & \multirow{3}{*}[-1.5ex]{\makecell[c]{Multimedia Learning\\Emotion Contagion}} & SP & \makecell[l]{segmenting, personalization, coherence, affect}  \\
\cmidrule{4-5}
& & & AC & \makecell[l]{coherence, signaling, voice naturalness, affect}  \\
\cmidrule{4-5}
& & & PE & \makecell[l]{coherence, spatial contiguity, signaling, affect}  \\
        \bottomrule
    \end{tabular}
    }
    \caption{The persuasion theories and applied psychological models used in the four stages of CogenPVG: argument reasoning (AR), storyboard planning (SP), asset creation (AC), and post-editing (PE).}
    \label{tab:criteria}
\end{table*}

To tackle the aforementioned challenge, this paper proposes a novel cognitive-enhanced reflective multi-agent framework specifically designed for the PVG task, referred to as CogenPVG. We choose to utilize the Elaboration Likelihood Model (ELM)~\cite{elm}, a well-established persuasion theory in psychology, as guidance in designing the framework. Specifically, the ELM defines two modes of persuasiveness perception: the central route and the peripheral route, corresponding to scenarios that involve deep thinking and superficial perception, respectively. Thus, we derive two principles for enhancing the persuasiveness of generated video: 1) conveying highly credible and logically rigorous information to enhance the central route, and 2) orchestrating content to utilize heuristics for peripheral route enhancement. As illustrated in Fig.~\ref{fig:teaser}, we follow these two principles as high-level guidance to construct our multi-agent framework, which consists of four sequential process stages: argument reasoning, storyboard planning, asset creation, and post-editing. In each stage, we employ a generator-critic reflective mechanism to optimize the generated content iteratively, ensuring that the outcome of each stage satisfies the persuasion criteria defined in the corresponding psychological models. The primary contributions of this paper could be summarized as follows:
\begin{itemize}
    \item We introduce a novel cognitive-enhanced reflective multi-agent framework, CogenPVG, for generating videos with high persuasiveness. To the best of our knowledge, CogenPVG is the first work on the PVG task for the visual persuasion on general topics.
    
    \item We incorporate the ELM persuasion theory and related applied psychological models throughout our framework by devising a generator-critic reflective mechanism with delicately designed prompts, ensuring the persuasiveness inherited from the persuasion theory.
    
    \item By conducting extensive experiments and evaluations on metrics that cover persuasiveness comparison, attitude shift, and a series of subjective dimensions, we validate the efficacy of our proposed framework and cognitively enhanced designs.
\end{itemize}

\section{Related Work}
\noindent\textbf{Persuasiveness Modeling.}
Modeling the persuasive effects and generating highly persuasive contents are significant research topics mainly in the computational linguistic domain. Some works~\cite{persuasive1, persuasion4good, persuasive_rewrite} model the persuasive strategies from public data and train language models to generate persuasive dialogue. Recently, some studies have shifted their focus to visual persuasion~\cite{pvg_survey, persuasive_storyline, persuasive2, cap, m2p2, imagearg, pvp}. CAP~\cite{cap} evaluates the persuasiveness of generated advertising images. M2P2~\cite{m2p2} predicts the persuasive outcome by analyzing the multimodal facial video. PVP~\cite{pvp} provides a large-scale image datasets with visual persuasiveness annotations and personal characteristics of the viewers.
Despite the advancements, the task of directly generate persuasive videos is still underexplored.

\noindent\textbf{LLM-based Video Generation.}
With the fast development of LLM models~\cite{gpt, gemini, deepseek}, recent works start to leverage their advanced semantic reasoning merits for video generation, especially in long narrative video cases~\cite{videoauteur, shen2025storygpt}. StoryGPT-V~\cite{shen2025storygpt} harnesses the context memorizing ability to address the ambiguous references, yielding better character consistency. VideoAuteur~\cite{videoauteur} proposes a LLM-augmented large-scale narrative video dataset and leverages a multimodal LLM as video director to acquire better visual consistency. Another approach is to leverage the subtask planning and tool calling capabilities in LLM via agentic frameworks~\cite{mmstory, movieagent, animaker, personavlog, vlogger, anim_director, aesopagent}. MM-StoryAgent~\cite{mmstory} simulates a discussing process for script planning, followed by a series of tool-calling agents for the assembly of multimodal videos. Vlogger~\cite{vlogger} hires a LLM as video director to generate vlog shooting plan and actor settings. Anim-Director~\cite{anim_director} employs a multimodal LLM to automatically orchestrate the entire animation-making process.
While extensive works have been conducted in the LLM-based video generation, a notable research gap persists in the PVG task, where the primary challenge is to change the subjective attitudes rather than improve the objective visual qualities.


\begin{figure*}[t]
    \centering
    \includegraphics[width=\linewidth]{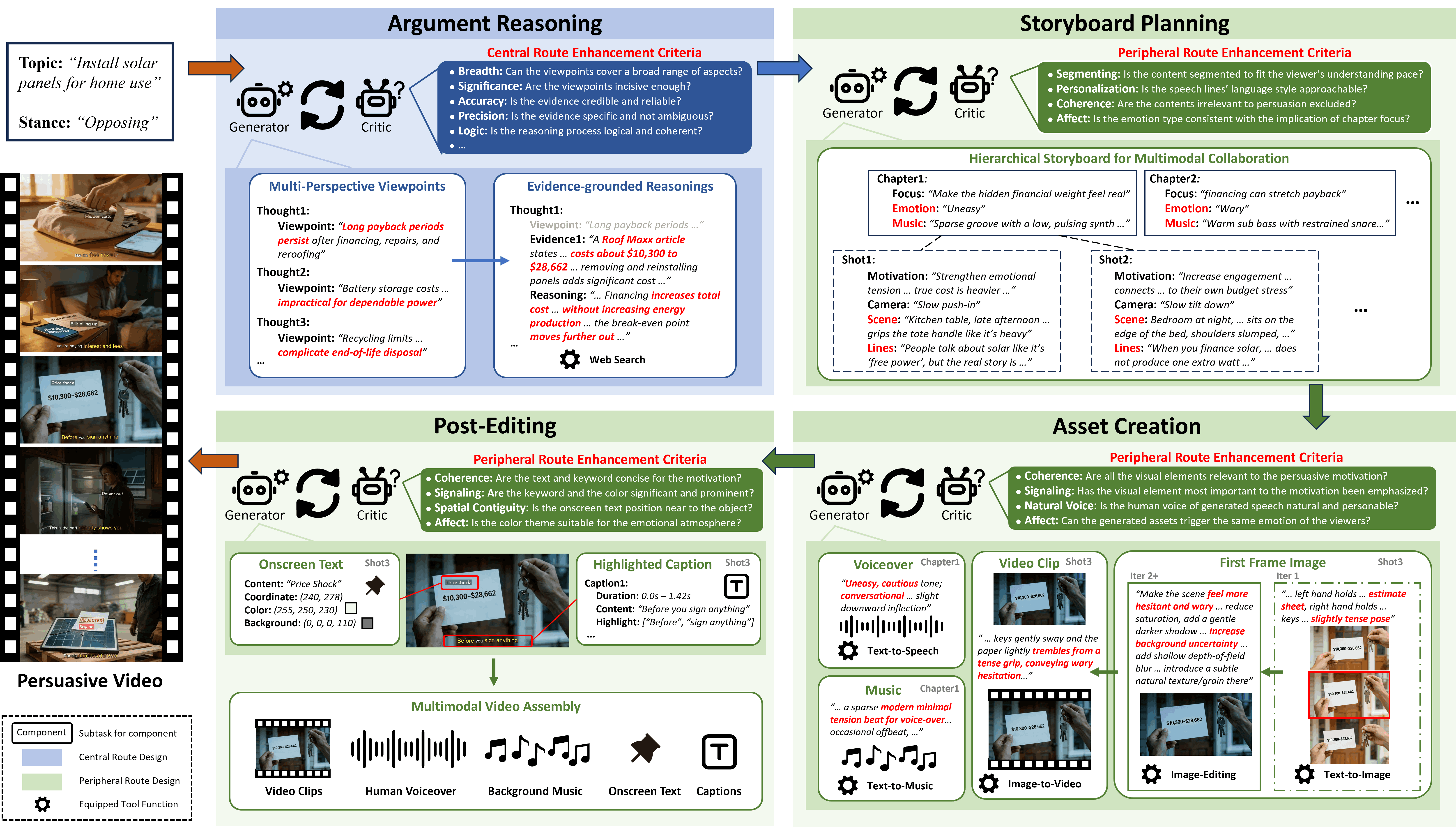}
    \caption{The framework of our proposed CogenPVG.
    The text highlighted in red illustrates the cognitive-guided generated content in the thinking quality, cognitive fluency, and emotional contagion.}
    \label{fig:pipeline}
\end{figure*}

\section{Methodology}

\subsection{Preliminary for ELM Theory}

The Elaboration Likelihood Model (ELM)~\cite{elm} is a dual-process model~\cite{dualprocess} of persuasion and attitude change~\cite{elm, elm1}, which suggests information can be processed in two distinct cognitive routes: the \textit{central route}, which involves high proactivity for careful thought of information credibility and logical soundness, and the \textit{peripheral route} relies on superficial cues with inferior cognitive resource requirement or elevated emotional arousal.
The ELM indicates two parallel aspects for modeling persuasiveness, where specific applied psychological models of \textit{Critical Thinking} theory~\cite{critical} and \textit{Heuristic} theory~\cite{heuristic} can be utilized for the central route and peripheral route, respectively.

Specifically, we adopt the representative and well-established Paul-Elder (P-E) model~\cite{pemodel} as the criteria instance of Critical Thinking in the central route. For the peripheral route, we focus on the \textit{fluency heuristic}~\cite{fluency, fluency2} and \textit{affect heuristic}~\cite{affectheuristic} which are instantiated as concrete creation guidelines based on their corresponding applied psychological models~\cite{multimedialearning, emotionalcontagion}. Please Refer to Appendix B for detailed explanations of psychological models and the criteria employed by the critic agents.

\subsection{Task Definition and Framework Overview}
\label{subsec:framework}
We define the persuasive video generation task as a mapping $PVG$ that projects a topic $t$ and a target stance $s$ to a highly persuasive video $V$,  as follows:
\begin{equation}
PVG: (t,s) \rightarrow V.
\end{equation}

The goal of generated video $V$ is to lead the audience to favor $s$ towards $t$. To this end, we incorporate the ELM persuasion theory and several practical psychological models as guidance in designing the four stages of our proposed framework, as illustrated in Fig.~\ref{fig:pipeline}. The four stages include {argument reasoning (AR)}, {storyboard planning (SP)}, {asset creation (AC)}, and {post-editing (PE)}, imitating the workflow of human video creators. 
In each stage, a pair of generator agent $G_\text{stage}$ and critic agent $C_\text{stage}$ execute one or multiple subtasks in an iterative and reflective refinement scheme, guaranteeing the effectiveness of ELM. 
As shown in Tab.~\ref{tab:criteria}, the argument reasoning stage generates content in the central route of persuasion, obeying the criteria of critical thinking, and the other three stages follow the peripheral route by instantiating the fluency and affect heuristics. Due to the page limit, we provide the detailed prompts for the generator and critic for each subtask in each stage in Appendix A.

\subsection{Argument Reasoning}
\label{subsec:ar}

Given the input topic $t$ and stance $s$, the argument reasoning stage aims to generate a set of high-quality thoughts $\mathcal{T}$ as persuasive content. Each thought consists of a viewpoint, several evidences, and a reasoning process generated by two subtasks: multi-perspective viewpoint generation and evidence-grounded reasoning.
\begin{equation}
    \mathcal{T} = {Stage}_\text{AR}(t, s).
\end{equation}

\noindent\textbf{Multi-Persepective Viewpoints Generation.}
To satisfy the \textit{breadth} criterion of the P-E model, we first prompt $G_\text{AR}$ to analyze the persuasion goal $(t,s)$ from diverse perspectives, including the benefits of the target stance and the potential drawbacks of the opposite stance.
After $G_\text{AR}$ forms the initial viewpoints, $C_\text{AR}$ reviews and critiques them based on more criteria of the P-E model. Specifically, $C_\text{AR}$ examines the \textit{relevance} between generated viewpoints and persuasion goal, the \textit{depth} of viewpoints to ensure a thorough exploration of the core issue, and the \textit{significance} of viewpoints to cover the most crucial aspects.

\noindent\textbf{Evidence-grounded Reasoning.} $G_\text{AR}$ conducts a two-step Chain-of-Thought (CoT)~\cite{cot} to collect evidential facts and generate reasoning based on the previously obtained viewpoints. We equip $G_\text{AR}$ with a web search tool and execute three queries for each viewpoint. Given the raw message returned from the queries, $G_\text{AR}$ extracts the critical information such as precise digits and authoritative institutions, in accordance with the \textit{accuracy} and \textit{precision} criteria of the P-E model. Then, it formulates a logically plausible reasoning based on the evidences to enrich each thought.
In $C_\text{AR}$, evidences and reasoning are reviewed simultaneously. $C_\text{AR}$ evaluates whether the evidences are sufficient to substantiate the corresponding viewpoint, while meeting the \textit{sufficiency} standard of the P-E model. For the reasoning, $C_\text{AR}$ inspects the \textit{logic} and \textit{clarity} to ensure that the generated thoughts $\mathcal{T}$ have no logical fallacies and are expressed unambiguously.

\subsection{Storyboard Planning}
\label{sbusec:sp}

Anchored in the argument thoughts $\mathcal{T}$ and the persuasion goal $(t,s)$ the storyboard planning generator ${G}_\text{SP}$ yields a hierarchical storyboard $\mathcal{S}$ in global chapter-level and precise shot-level via CoT prompting.
\begin{equation}
\mathcal{S} = {Stage}_\text{SP}(t, s, \mathcal{T}).
\end{equation}

In chapter-level planning, $G_\text{SP}$ semantically clusters the thoughts $\mathcal{T}$ into chapters based on different focused issues, aiming to decrease cognitive load by obeying the \textit{segmenting} principle in the Multimedia Learning theory. Then, to leverage the affect heuristic, $G_\text{SP}$ assigns each chapter a proper emotion and articulates a music description for each chapter.
In shot-level planning, $G_\text{SP}$ further divides each chapter into multiple shots. Each shot consists of a concrete motivation corresponding to $\mathcal{T}$, $t$, and $s$, as well as descriptions of the visual scene, camera trajectory, and speech lines. 
To ameliorate, $C_\text{SP}$ evaluates all chapters and shots in a single prompting step, enabling the joint optimization of both levels. $C_\text{SP}$ first assesses the chapter arrangement and the distinctiveness among different chapters to facilitate comprehension, as proposed in the \textit{segmenting} criterion. Then it judges the rationality of the emotional setting and guarantees that the music description can effectively convey the target emotion. For the shot-level optimization, $C_\text{SP}$ sticks to the \textit{personalization} and \textit{coherence} criteria to evaluate whether the speech lines exhibit an approachable language style and contribute to the shot motivation, as well as the emotional grounding of visual scene description.

\subsection{Asset Creation}
Following the chapter and shot descriptions in $\mathcal{S}$, the agents generate and optimize a range of multimodal video assets by three subtasks: video clip $\mathcal{V}$, human voiceover clip $\mathcal{H}$, and background music clip $\mathcal{M}$ generation.
\begin{equation}
    \{\mathcal{V}, \mathcal{H}, \mathcal{M}\} = {Stage}_\text{AC}(\mathcal{S}).
\end{equation}

\noindent\textbf{Video Clip Generation.}
Instead of generating video clips directly via text-to-video models, we use images as an intermediate modality to fully leverage image-editing capabilities. Firstly, we prompt ${G}_\text{AC}$ to call the text-to-image tool with an optimized emotion-centered prompt and generate several first-frame image candidates. $G_\text{AC}$ then selects the best one with the fewest artifacts. The generated image is reviewed by $C_\text{AC}$ based on the \textit{coherence} and \textit{signaling} criteria in the Multimedia Learning theory, which suggests the image to contain only necessary elements relevant to the shot motivation and emphasize the core visual element that powerfully conveys the motivation. It also provides feedback on the emotional properties of the image based on the \textit{affect} criteria, as outlined in the Emotional Contagion theory. Notably, the reflective iteration in first-frame image generation is achieved by switching to the image-editing tool, which directly optimizes the visual contents in image scope rather than prompt scope. Then, $G_\text{AC}$ creates video clips using the image-to-video tool.

\noindent\textbf{Voiceover and Music Clip Generation.}
We directly utilize the speech lines in $\mathcal{S}$ as the voiceover scripts, while prompting $G_\text{AC}$ with the emotion to infer emotion-centered instructions regarding the speaking tone.
Following the same process of rewriting the description to an emotion-centered prompt and generating assets by calling generation tool, $G_\text{AC}$ creates the chapter-level background music clips.
As the primary objective of auditory assets is to amplify the emotional atmosphere, we adopt the \textit{affect} criteria in $C_\text{AC}$ for the subtasks of voiceover and music clip generation to enhance emotional resonance.

\subsection{Post-Editing}
\label{subsec:pe}
After acquiring the multimodal assets, the post-editing stage aims to enhance the persuasiveness of the final video through two subtasks: onscreen text $\mathcal{R}$ and highlighted caption $\mathcal{C}$ generation, and then assembles all the video components into a complete multimodal persuasive video $V$.
\begin{equation}
    \{\mathcal{R}, \mathcal{C}\} = {Stage}_\text{PE}(\mathcal{V}, \mathcal{H}).
\end{equation}

\noindent\textbf{Onscreen Text Generation.}
According to the \textit{signaling} criterion, we prompt $G_\text{PE}$ with the first frame image and shot motivation to synthesize an onscreen text with attributes of font color, background color, and pixel coordinates.
Based on the \textit{coherence} and \textit{spatial contiguity} criteria, $C_\text{PE}$  provides recommendations to convey the core motivation within the text content concisely and to position it adjacent to the relevant visual element. Regarding the \textit{signaling} and \textit{affect} criteria, color settings are adjusted to enhance text readability and to harmonize the emotional tone with the chapter's overall sentiment.

\begin{table*}[t]
\centering
{\fontsize{8}{\baselineskip}\selectfont
\setlength{\tabcolsep}{1mm}

\begin{tabular}{lllllllllllll}
\toprule
\multirow{2}{*}{Method} & \multicolumn{3}{l}{Persuasiveness (Human)} & \multicolumn{3}{l}{Persuasiveness (LMM)}  & \multicolumn{6}{l}{Attitude Shift} \\
                        & Win ($\uparrow$)        & Tie         & Loss ($\downarrow$)        & Win ($\uparrow$)        & Tie          & Loss ($\downarrow$)        & AAS ($\uparrow$) & AS$_1$ ($\uparrow$) & AS$_2$ ($\uparrow$) & AS$_3$ ($\uparrow$) & AS$_4$ ($\uparrow$) & AS$_5$ ($\uparrow$)\\
\midrule
MM-StoryAgent           & \textbf{0.636}     & 0.143     & 0.221     & \textbf{0.653}     & 0.042    & 0.305    & 0.243          & 2.648          & 1.944          & 1.225          & 0.676 & -0.254 \\
Anim-Director           & \textbf{0.788}     & 0.120     & 0.092     & \textbf{0.901}     & 0.056    & 0.043    & 0.127          & 2.606          & 1.873          & 1.099          & 0.732 & -0.254 \\
VideoGen                & \textbf{0.688}     & 0.118     & 0.194     & \textbf{0.915}     & 0.042    & 0.043    & 0.257          & 2.451          & 1.802          & 1.014          & 0.690 & -0.155 \\
DirectPVG               & \textbf{0.541}     & 0.264     & 0.195     & \textbf{0.662}     & 0.000    & 0.338    & 0.347          & \textit{3.197} & \textit{2.229} & 1.443          & \textbf{1.000} & \textbf{0.000}  \\
AutoPVG                 & \textbf{0.493}     & 0.285     & 0.222     & \textbf{0.606}     & 0.014    & 0.380    & \textit{0.444} & 3.127          & 2.142          & \textit{1.451} & 0.930 & -0.042 \\
CogenPVG                & -                  & -         & -         & -                  & -        & -        & \textbf{0.604} & \textbf{3.268} & \textbf{2.296} & \textbf{1.592} & \textbf{1.000} & \textbf{0.000}   \\
\bottomrule

\end{tabular}
}
\caption{Quantitive results of all methods. Our CogenPVG consistently prevails over baseline methods, with a win rate notably higher than the loss rate across every comparison setting and the largest positive attitude shift toward the persuasion goal.}
\label{tab:compare}
\end{table*}

\begin{table*}[t]
\centering
{\fontsize{8}{\baselineskip}\selectfont
\setlength{\tabcolsep}{1mm}
\begin{tabular}{lcccccccccccccc}
\toprule
\multicolumn{1}{c}{\multirow{2}{*}{Method}} & \multicolumn{2}{c}{Persu}     & \multicolumn{2}{c}{Arg}       & \multicolumn{2}{c}{Cred}      & \multicolumn{2}{c}{Aff}       & \multicolumn{2}{c}{Comp}      & \multicolumn{2}{c}{Nat}      & \multicolumn{2}{c}{Mem}       \\
\multicolumn{1}{c}{} & M & \textbf{H} & M & \textbf{H} & M & \textbf{H} & M & \textbf{H} & M & \textbf{H} & M & \textbf{H} & M & \textbf{H}\\
\midrule
w/o both       & 4.667          & \textit{3.569} & 4.639          & \textit{3.327} & \textit{3.028} & 2.957          & 4.361          & 3.020          & \textbf{4.986} & \textit{3.783} & \textbf{4.944} & \textit{3.652} & \textit{4.028} & 2.980          \\
w/o central    & \textit{4.750} & 3.417          & 4.618          & 3.245          & 2.882          & 3.060          & \textbf{4.500} & \textit{3.264} & 4.944          & 3.740          & 4.861          & 3.560          & \textbf{4.056} & \textit{3.038} \\
w/o peripheral & 4.676          & 3.167          & \textit{4.694} & 3.241          & 2.917          & \textit{3.135} & \textit{4.444} & 2.981          & 4.917          & 3.596          & 4.861          & 3.481          & \textit{4.028} & 2.852          \\
Ours                & \textbf{4.801} & \textbf{3.764} & \textbf{4.722} & \textbf{3.542} & \textbf{3.083} & \textbf{3.278} & 4.382          & \textbf{3.556} & \textit{4.971} & \textbf{4.069} & \textit{4.882} & \textbf{3.944} & 4.000          & \textbf{3.431}  
  \\
\bottomrule
\end{tabular}
}
\caption{Quantitative results of the ablation study, containing both LMM-based (M) and human-based (H) rating results.
}
\label{tab:abl}
\end{table*}

\begin{figure}[t]
    \centering
    \includegraphics[width=\linewidth]{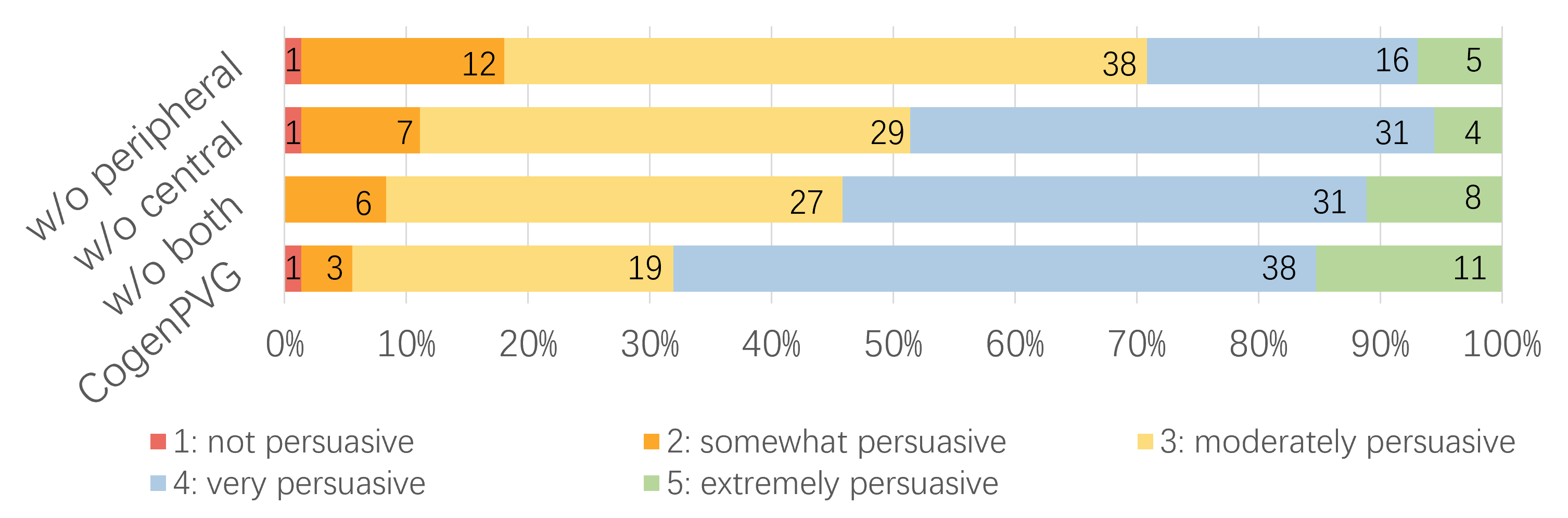}
    \caption{Distribution of human-based persuasiveness ratings in ablation study.}
    \label{fig:likert}
\end{figure}

\begin{figure*}[t]
    \centering
    \includegraphics[width=0.9\linewidth]{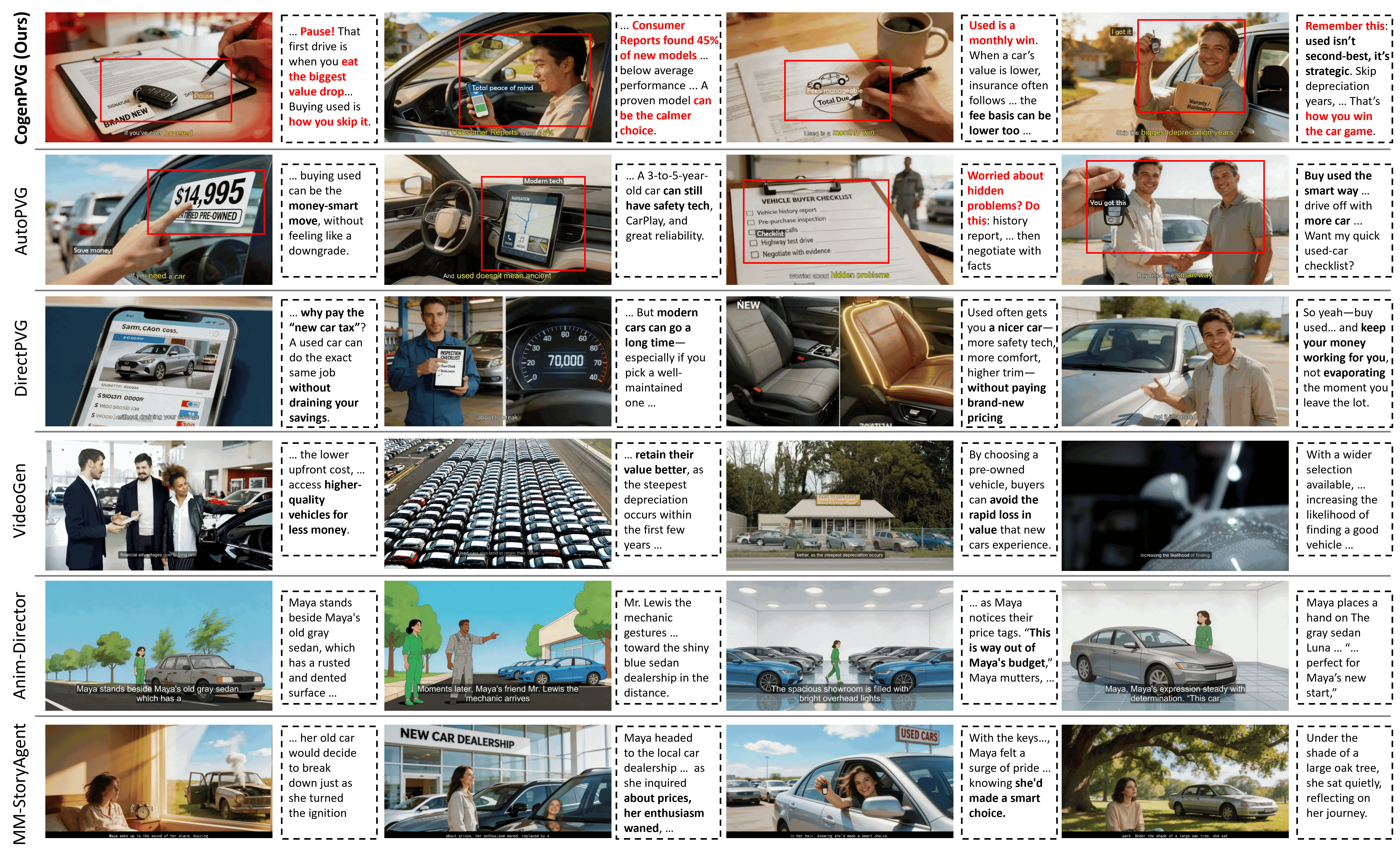}
    \caption{Qualitative results. We present the generated video snapshots and the corresponding speech lines on the topic \textit{Buy a used car} with stance \textit{supporting} as the persuasion goal. The speech contents that contribute to the persuasion goal are rendered in bold, and the ELM-related cognitive-enhanced contents are highlighted in red. The visual centers that closely serve the persuasion goal are marked with red boxes.}
    \label{fig:vis}
\end{figure*}

\noindent\textbf{Highlighted Caption Generation.}
To explicitly reduce the cognitive load, $G_\text{PE}$ segments the speech lines into captions where key elements, such as precise statistical numbers and emotive words, are highlighted in different colors and fonts.
$C_\text{PE}$ critiques the captions $\mathcal{C}$ according to the criteria of \textit{coherence} and \textit{signaling} to optimize the selection of the most concise and significant words.

After all the video components $\mathcal V, \mathcal H, \mathcal M, \mathcal{R}, \mathcal{C}$ are generated and optimized, we combine shot-level components, and apply a smooth fading transition between chapters for the video and audio channel, to form the final persuasive video $V$.
    \begin{equation}
        V = {Assemble}(\mathcal{V}, \mathcal{H}, \mathcal{M}, \mathcal{R}, \mathcal{C}).
    \end{equation}

\section{Experiment}
\noindent\textbf{Data Preparation.} 
Since there is no conventional benchmark for the PVG task, we construct the input topic and stance set from the Personalized Visual Persuasion (PVP) dataset~\cite{pvp}. 
After removing two potentially sensitive themes (\textit{defense} and \textit{veteran affairs}), we retain 18 themes and select two messages per theme to form our topic set. For each topic, we then assign two stances (i.e., supporting and opposing), resulting in 72 persuasion goals in total. Given that persuasive generation task is inherently more subjective than aesthetic generation task, we construct an evaluation set several times larger than those used in other story visualization works, where about 10 samples are typically considered sufficient to draw solid conclusions~\cite{animaker, vlogger, movieagent, autostory}.

\noindent\textbf{Implementation Details.}
We utilize GPT-5.4 for the generator agents and GPT-4o for the critic agents, which are widely regarded as top-tier large multimodal models (LMMs).
Given that GPT-4o API lacks native audio processing ability, we additionally employ Qwen3-Omni~\cite{xu2025qwen3_omni} to analyze voiceover and music clips. For the equipped tools, we employ Tavily Search~\cite{tavily} for evidence searching in $G_\text{AR}$, Seedream-4.5~\cite{seedream} for image generation and editing, Seedance-1.5-pro~\cite{seedance}, Seed-tts-2.0~\cite{seed_tts}, and Suno-4.5 for the generation of video clips, human voiceover and background music clips in $G_\text{AC}$ respectively. All tools are wrapped as MCP services 
and proactively invoked by the agents. 
Please refer to Appendix A for our detailed implementations, including the system prompts and subtask executions.

\noindent\textbf{Baselines.}
Because CogenPVG is the first framework for persuasive video generation on general topics, it is hard to find existing methods strictly aligned with our task setting for a fully fair comparison. Therefore, we select three of the most relevant agentic long-form video synthesis methods and make minor adaptations for comparative discussion. Specifically, we use 1) \textbf{MM-StoryAgent}~\cite{mmstory} and 2) \textbf{Anim-Director}~\cite{anim_director} as representative narrative-centered video generation methods for comparison, whose input synopses are generated by prompting GPT-5.4\cite{openai_gpt54} with our persuasion goals. We further include 3) \textbf{VideoGen}~\cite{videogen_platform}, a prevailing commercial platform that synthesizes videos by retrieving contextually relevant clips from a vast stock media library, as a representative industrial baseline.
To better evaluate our method, we propose two degraded versions of our method as additional comparison baselines, where 4) \textbf{DirectAgent} removes all cognitive-enhanced designs and retains only the generators in the last three processing stages, serving as a minimal video synthesis framework, and 5) \textbf{AutoPVG} consolidates the four stages into a single agent that acomplishes the PVG task by autonomously planning and tool execution.

\noindent\textbf{Metrics.}
Following ~\cite{enhancingpersu, mmpersuade, persugpt}, we evaluate our method and the selected baselines via Win-Tie-Loss persuasiveness comparison and attitude shift scores. Both categories of metrics synergize human-based user studies and LMM-based automated assessments. Human evaluation serves as the gold standard for real-world persuasive efficacy, whereas 
LMM is exploratorily employed as a potential scalable and objective estimation that mitigates the influence of subjective individual biases.
Attitude is quantified using a 5-point Likert scale, ranging from 1 (strongly disagree) to 5 (strongly agree). Average Attitude Shift (AAS) denotes the average increase in human participants' attitudes. 
AS$_i$ measures the increase of attitude value from initial score $i$. Since human participants' pre-existing attitudes are non-manipulable, we utilize LMM with predefined initial states for AS$_i$ to estimate the persuasive efficacy with different initial attitudes. To alleviate self-preference bias~\cite{selfprefer},
we employ Gemini-2.5-pro as the automatic evaluator, a model distinct from the LMMs applied in our framework. It is instructed to act as a lay audience without psychology expertise and provide first-person reasoning for the assessment results.

To further analyze how different cognitive-enhancement routes affect persuasive outcomes, we additionally adopt a set of subjective rating metrics in the ablation study, using five-point Likert scale to quantify multidimensional influencing factors. Similar to related cognitive research~\cite{heuristic, elm1}, we include overall persuasiveness (Persu), argument quality (Arg), credibility (Cred), affective potency (Aff), comprehension fluency (Comp), perception naturalness (Nat), and memorability (Mem). 
Please refer to the Appendix C for their definitions and evaluation prompts.


\noindent\textbf{Baseline Comparison.}
For pairwise persuasion comparison, we combine videos generated by five baseline methods with those generated by CogenPVG to construct 360 video pairs. 
Each video pair is evaluated twice in random order by 36 human participants, yielding a total of 720 evaluation samples. The Cohen's $\kappa$ between the two evaluation rounds is 0.325, which is typical for subjective tasks~\cite{mmpersuade}. For LMM-based automatic evaluation, we swap the internal order across two evaluation rounds to mitigate position bias~\cite{positionbias}.
A similar protocol is adopted for attitude shift evaluation on 432 generated videos, yielding 864 attitude shift samples in the human-based evaluation and 864 samples per initial LMM attitude score.

Tab.~\ref{tab:compare} presents the quantitative results, where win-tie-loss ratios are computed by comparing our CogenPVG against baseline methods. Results demonstrate that our method consistently achieves a higher win rate than loss rate, underscoring the superior persuasiveness of CogenPVG. Although DirectPVG and AutoPVG are degraded variants, they still preserve a design tailored exclusively to the nuances of persuasive video generation, therefore show comparatively lower win rate and higher tie rate than other baselines. We observe that LMM shows a pronounced preference for our method over Anim-Director, indicating that LMM tends to disfavor animated visual styles for persuasion. A similar LMM preference occurs in the comparison against VideoGen, which can be attributed to the inferior alignment between its visual synthesis and vocal commentary. 
While the magnitude of preference differs, human and LMM-based evaluations convergently indicate that our method outperforms the baselines in the pairwise persuasiveness comparison. 

Our analysis of attitude shift in LMM reveals a ceiling effect: as initial attitude scores ascend, the marginal efficacy of persuasive in further augmenting endorsement diminishes. Notably, baseline methods often struggle to consolidate existing support and may even trigger adverse outcomes. Nevertheless, our method along with its degraded variants remains highly effective in eliciting attitude shifts towards the ceiling. The highest average attitude shift in human-based evaluation further demonstrates the effectiveness. 

\noindent\textbf{Ablation Study.}
To better understand our generator-critic reflective mechanism, 
we conduct three ablation experiments on 36 persuasion goals through removing the critic agent in the central route (i.e., $C_\text{AR}$), the peripheral route (i.e., $C_\text{SP}, C_\text{AC}, C_\text{PE}$), and both routes, forming \textit{w/o central}, \textit{w/o peripheral} and \textit{w/o both} respectively. 
The ablated results are shown in Tab.~\ref{tab:abl}.
1) \textit{w/o both}: 
The framework tends to generate natural and fluent content when no critic agent is involved, but remains less compelling in emotional arousal and evidential credibility, yielding a suboptimal persuasive effect.
2) \textit{w/o central}: The generated videos exhibit enhanced emotional expressiveness, albeit at the cost of credibility. 
3) \textit{w/o peripheral}: The results show a noticeable increase in the content credibility, but the poor fluency and affective potency performance indicates that the generated videos are difficult to follow and memorize with light cognitive resources, resulting in lower persuasiveness. 
Interestingly, we observe that the two ELM routes do not contribute to persuasion in a cumulative manner. Instead of incremental gains, the independent enhancement of single route appears to undermine overall efficacy. This implies that persuasion depends on a synergistic balance, where over-emphasizing one route may trigger cognitive interference in some subjective dimensions, thus lead to diminished persuasiveness.


Through grounding the persuasion in more precise and credible information collected from factual evidence and evoking higher affective potency together, CogenPVG 
achieves superior persuasive efficacy, as validated by both human-based evaluations and LMM-based estimations. Additionally, we present the distribution of human ratings in Fig.~\ref{fig:likert}, identifying that our method yields persuasiveness scores that are more concentrated in the high-score range.

\noindent\textbf{Qualitative Analysis.}
We present the generation results in Fig.~\ref{fig:vis}. MM-StoryAgent and Anim-Director converts the persuasion goal into story-based character settings and compelling plots.
These methods prioritize story progression over viewpoint delivery, thereby diluting persuasive impact. In addition, their lack of persuasion sensitivity can lead to the omission of goal-critical content: Anim-Director overlooks the explicitly stated new-versus-used car comparison in the synopsis and concludes merely by recommending the purchase of a favored car.
We attribute this phenomenon to the intrinsic difference between narrative stories and persuasive scripts, which reveals the notable gap between existing video synthesis tasks and PVG task, validating the necessity of our work.
VideoGen adopts a relatively direct presentation strategy with limited viewpoint coverage, relying mainly on cost-saving arguments in an explanatory persuasion style, which weakens its overall persuasive efficacy.
Benefiting from its structured script orchestration process, DirectPVG demonstrates stronger adherence to the persuasion goal.
However, without the dual-route enhancements of ELM, its argumentation remains relatively shallow and weakly grounded in factual evidence. Furthermore, it fails to establish information saliency.
By incorporating ELM principles, AutoPVG achieves better cognitive load control and attention guidance through visual cues, thereby yielding stronger persuasiveness. Nevertheless, in the absence of critic agents, its cognitive-enhancement gains remain limited.

By decoupling the workflow into four stage-specialized agents and applying ELM-guided generator–critic reflective refinement at each stage, our CogenPVG achieves stronger emotion leading, higher argument density, and more effective information delivery.
The cognitive enhancements enable our framework to outperform the two degraded baselines and achieve the best persuasiveness performance.

\noindent\textbf{Discussion and Limitations.} 
Our framework leverages pre-trained, off-the-shelf generative models for each modality, enabling convenient integration of latest models (Appendix D.1). However, it introduces limitations in terms of cost and computational efficacy (Append D.2).
We have also probed the viability of LMMs as neutral evaluators, free from the subjective idiosyncrasies of individual human raters. Nevertheless, their role in assessing video persuasion warrants granular investigation. 

\section{Conclusion and Future work}
\label{sec:conclusion}
This paper presents CogenPVG, a cognitive-enhanced reflective multi-agent framework specifically tailored for the persuasive video generation task, expanding persuasive generation research from linguistic domain into multimodal videos. Inspired by the well-established ELM persuasion theory, we 
govern our framework with two primary principles: the central route for high-quality rational reasoning, and the peripheral route for multimodal cues that facilitate mental shortcuts. By incorporating a generator-critic reflective mechanism, CogenPVG effectively amplifies the persuasiveness rooted in solid persuasion psychological models. Extensive experiments demonstrate that our framework effectively synergizes the dual routes, resulting in highly persuasive video outputs.

Our near-term research goal is to extend our framework with RAG capability, 
further enhancing the fluidity and credibility by retrieving real-world content.
We regard the development for other persuasiveness factors, such as personal traits and persuasion strategies, as a future long-term work.

\bibliography{aaai2027}


\appendix

\section{A Implementation Details}
\label{supp:impl}


\subsection{System Prompts}
\label{supp:prompt}

\begin{figure*}[h]
    \centering
    \begin{subfigure}[b]{\linewidth}
        \centering
        \includegraphics[width=\linewidth]{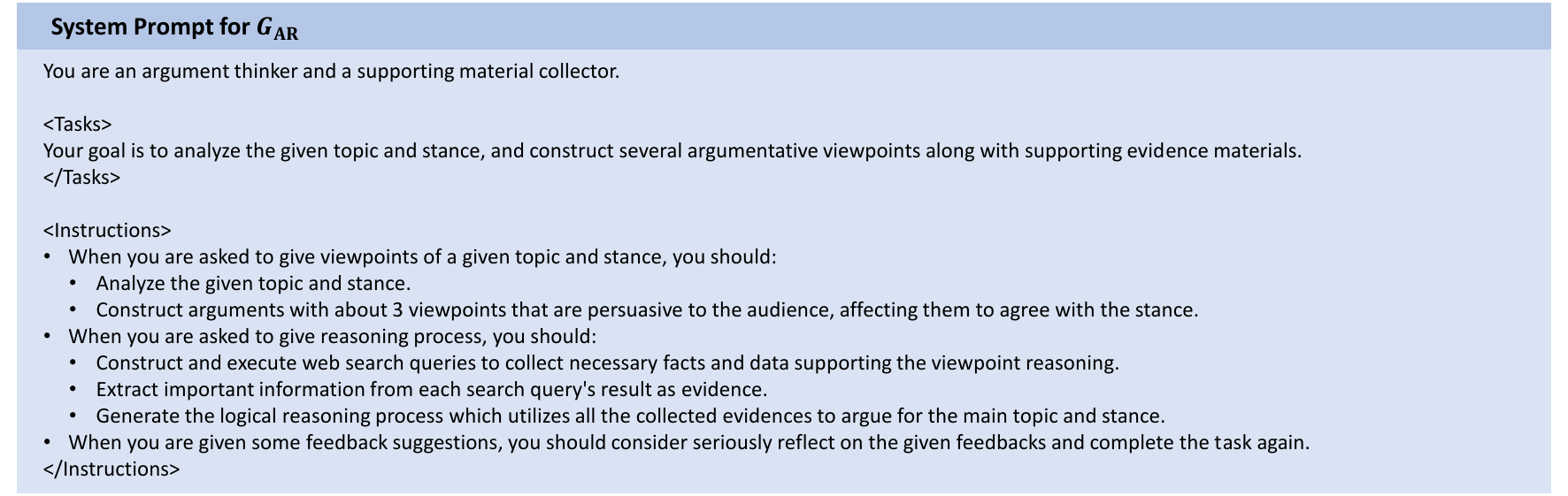}
    \end{subfigure}
    \hfill
    \begin{subfigure}[b]{\linewidth}
        \centering
        \includegraphics[width=\linewidth]{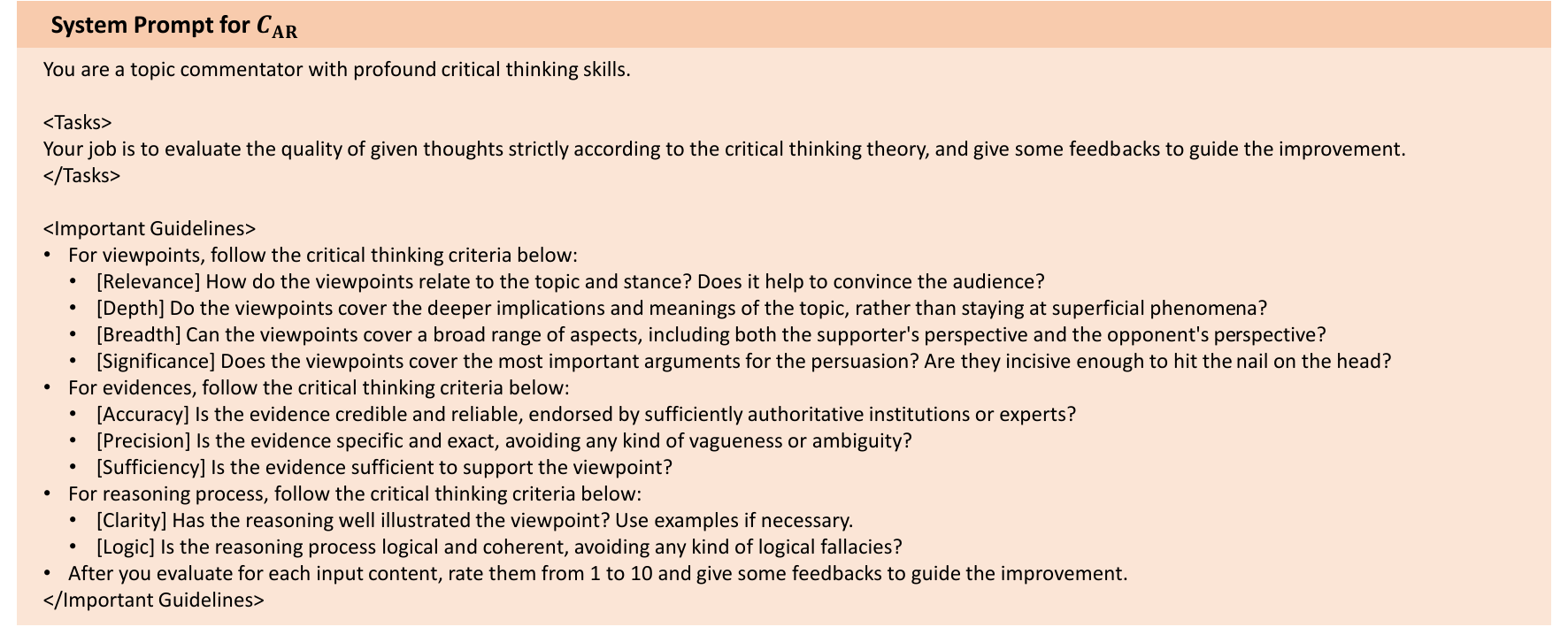}
    \end{subfigure}

    \caption{System prompts in argument reasoning stage.}
    \label{fig:prompt_ar}
\end{figure*}

\begin{figure*}
    \centering
    \begin{subfigure}[b]{\linewidth}
        \centering
        \includegraphics[width=\linewidth]{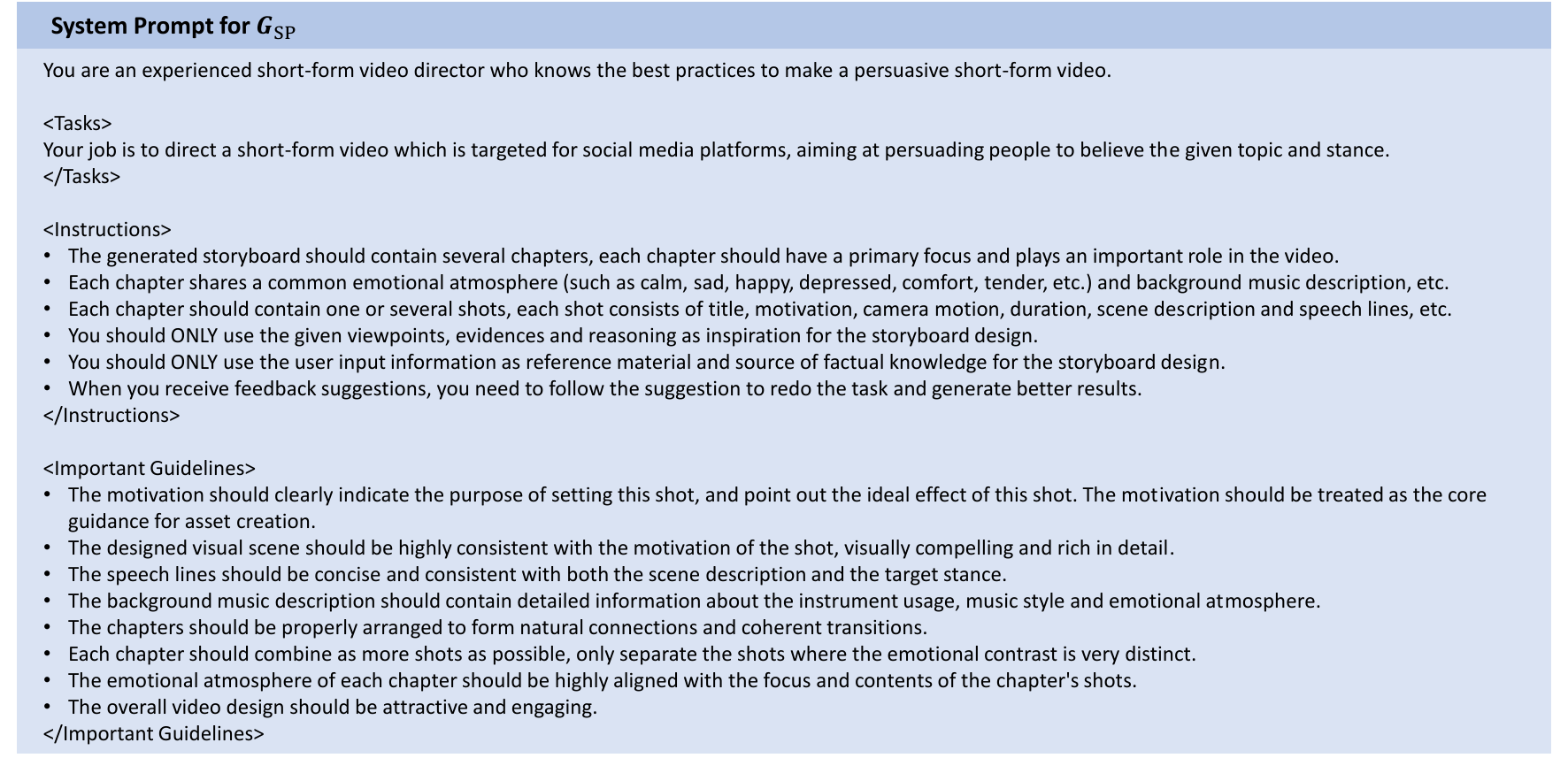}
    \end{subfigure}
    \begin{subfigure}[b]{\linewidth}
        \centering
        \includegraphics[width=\linewidth]{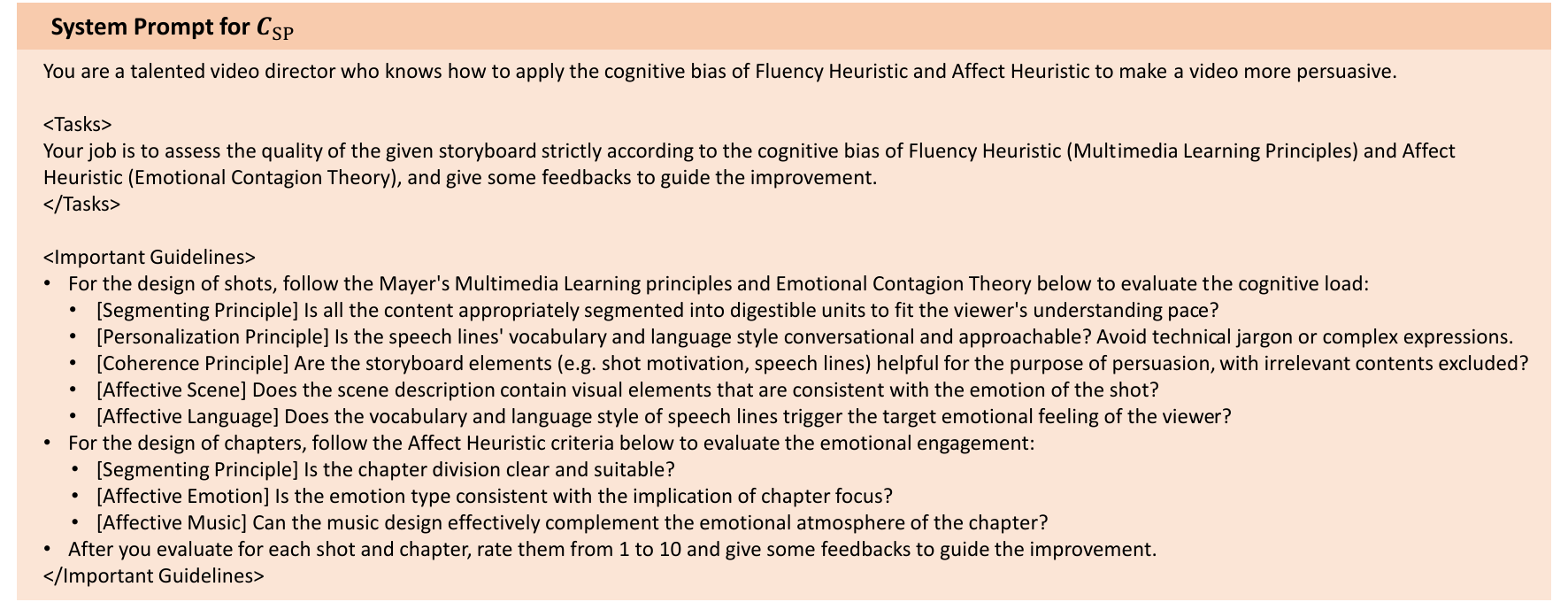}
    \end{subfigure}

    \caption{System prompts in storyboard planning stage.}
    \label{fig:prompt_sp}
\end{figure*}

\begin{figure*}
    \centering
     \begin{subfigure}[b]{\linewidth}
        \centering
        \includegraphics[width=\linewidth]{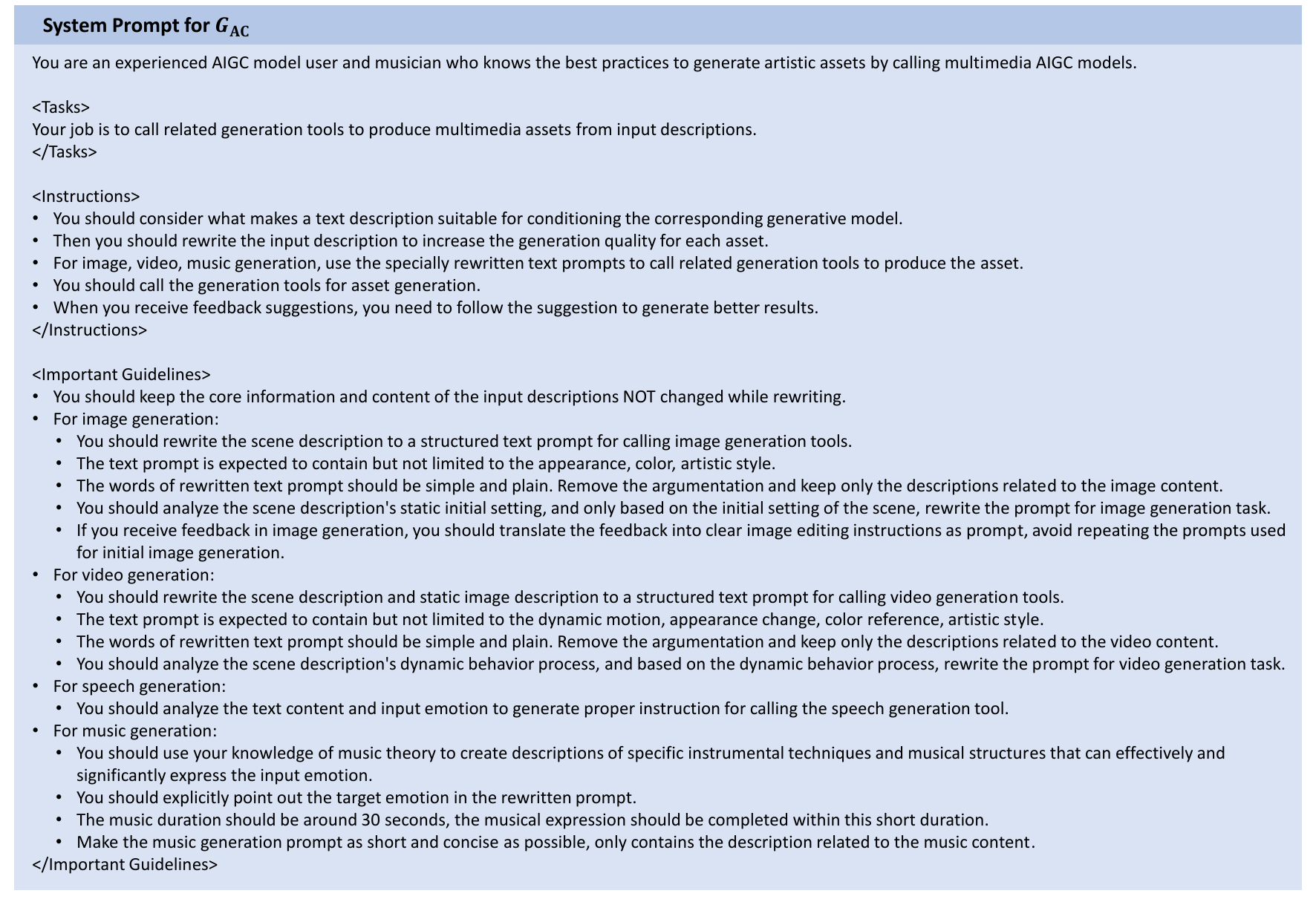}
    \end{subfigure}
    \hfill
    \begin{subfigure}[b]{\linewidth}
        \centering
        \includegraphics[width=\linewidth]{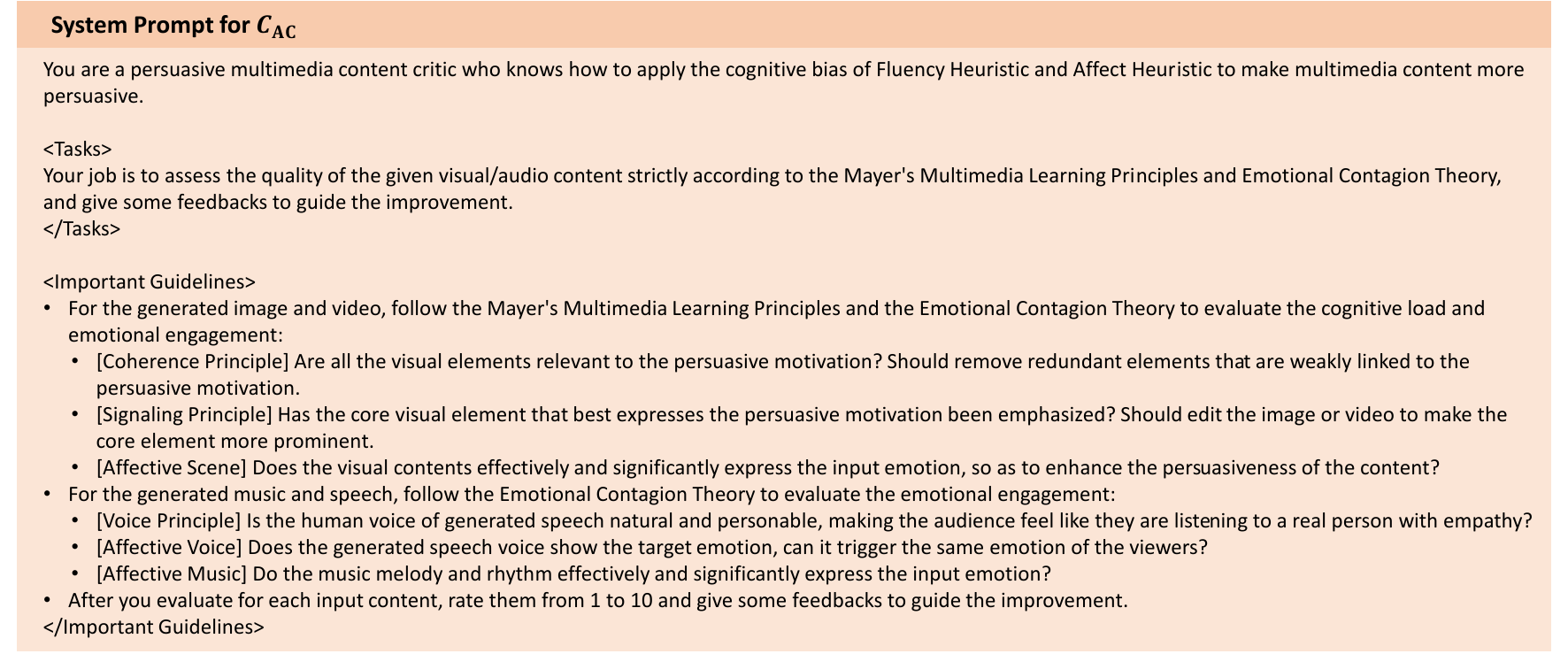}
    \end{subfigure}
    
    \caption{System prompts in asset creation stage.}
    \label{fig:prompt_ac}
\end{figure*}

\begin{figure*}
    \centering
     
    \begin{subfigure}[b]{\linewidth}
        \centering
        \includegraphics[width=\linewidth]{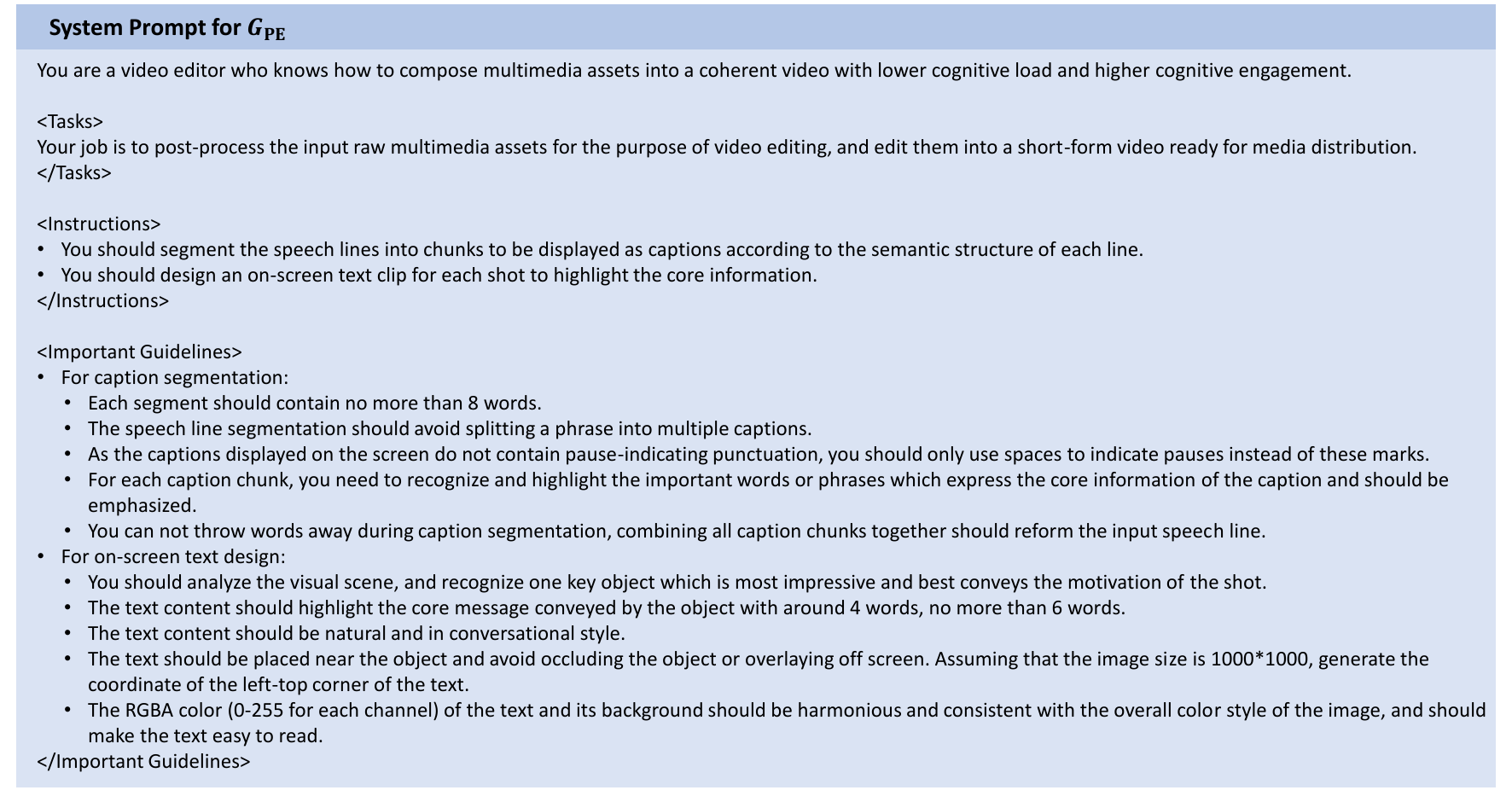}
    \end{subfigure}
    \hfill
    \begin{subfigure}[b]{\linewidth}
        \centering
        \includegraphics[width=\linewidth]{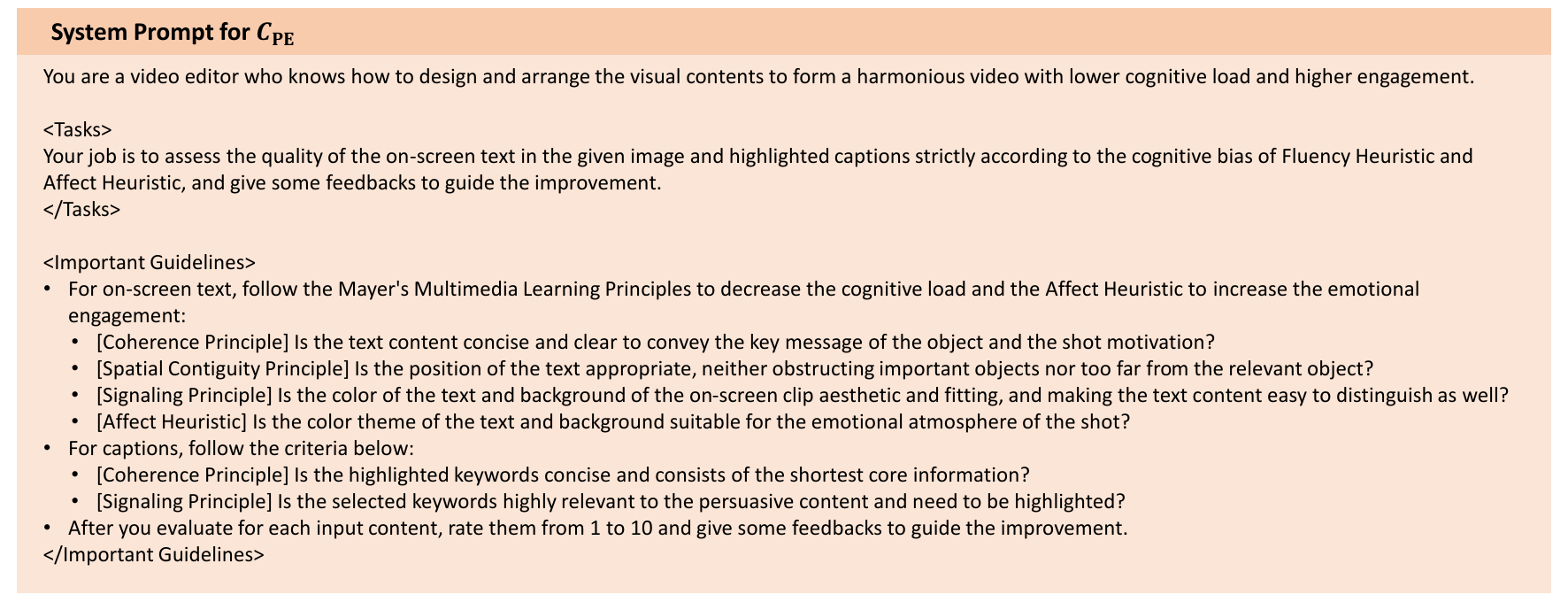}
    \end{subfigure}
    
    \caption{System prompts in post-editing stage.}
    \label{fig:prompt_pe}
\end{figure*}

We implement the generator and critic agents in our framework by contextual prompting where the cognitive-enhanced functional specialization of agents is defined by their respective dedicated system prompts. We display the main content of these system prompts in Fig.~\ref{fig:prompt_ar}, \ref{fig:prompt_sp}, \ref{fig:prompt_ac} and ~\ref{fig:prompt_pe}.
In the beginning of each system prompt, the main role of the agent is defined by one-sentence description, followed by a series of tagged regions, where \textit{\textless Tasks\textgreater} encapsulates the primary objective, \textit{\textless Instructions\textgreater} lists the execution chain and describes the basic requirements for each subtask, and \textit{\textless Important Guidelines\textgreater} specifies the detailed constraints for generator agents and related critiquing principles for critic agents.
\subsection{Subtask Executions}
\begin{figure*}
    \centering
     \includegraphics[width=\linewidth]{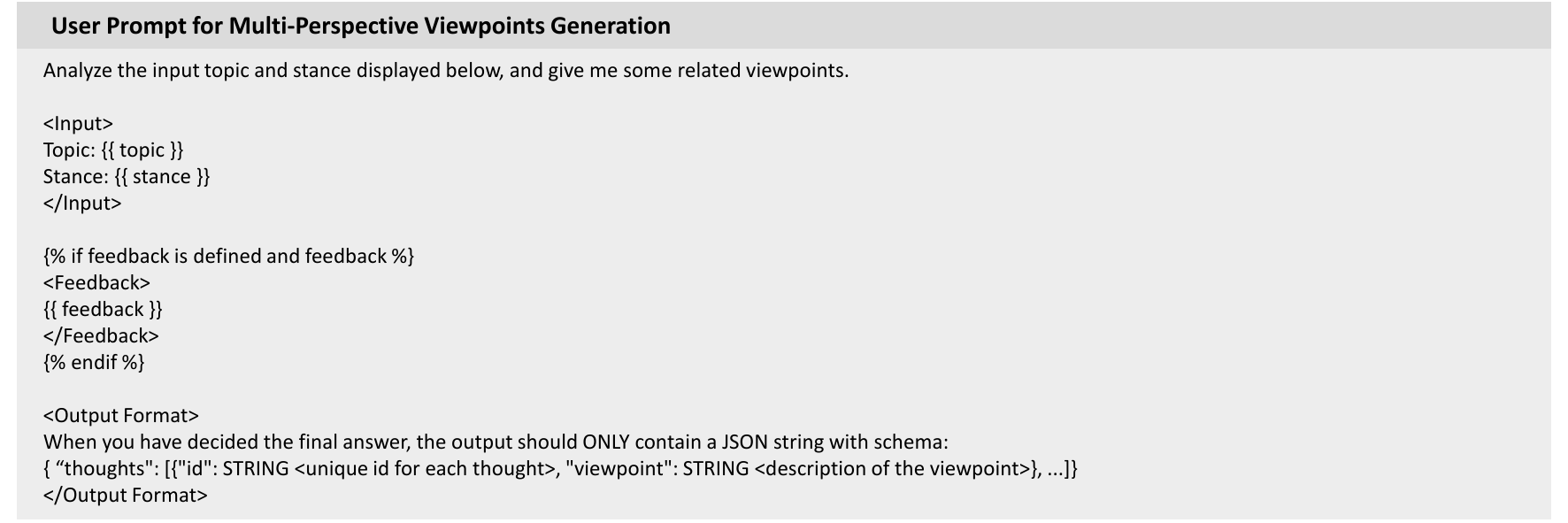}
    \caption{User prompt for triggering viewpoint generation subtask.}
    \label{fig:prompt_user}
\end{figure*}

In each processing stage, we execute the subtasks by sending a user prompt that contains the necessary input arguments to the agents, and receiving the json-structured response that can be parsed to extract the output results of the subtask.
The dynamic conversational context of each subtask is maintained and utilized in the reflective iteration to avoid ineffective update.
After the reflection converges, next subtask is executed samely.
We display the user prompt template for multi-perspective viewpoints generation subtask in Fig.~\ref{fig:prompt_user}.
For more details on the user prompts designed to trigger subtasks, please refer to our code files in the supplement.

\begin{figure*}
    \centering
    \includegraphics[width=\linewidth]{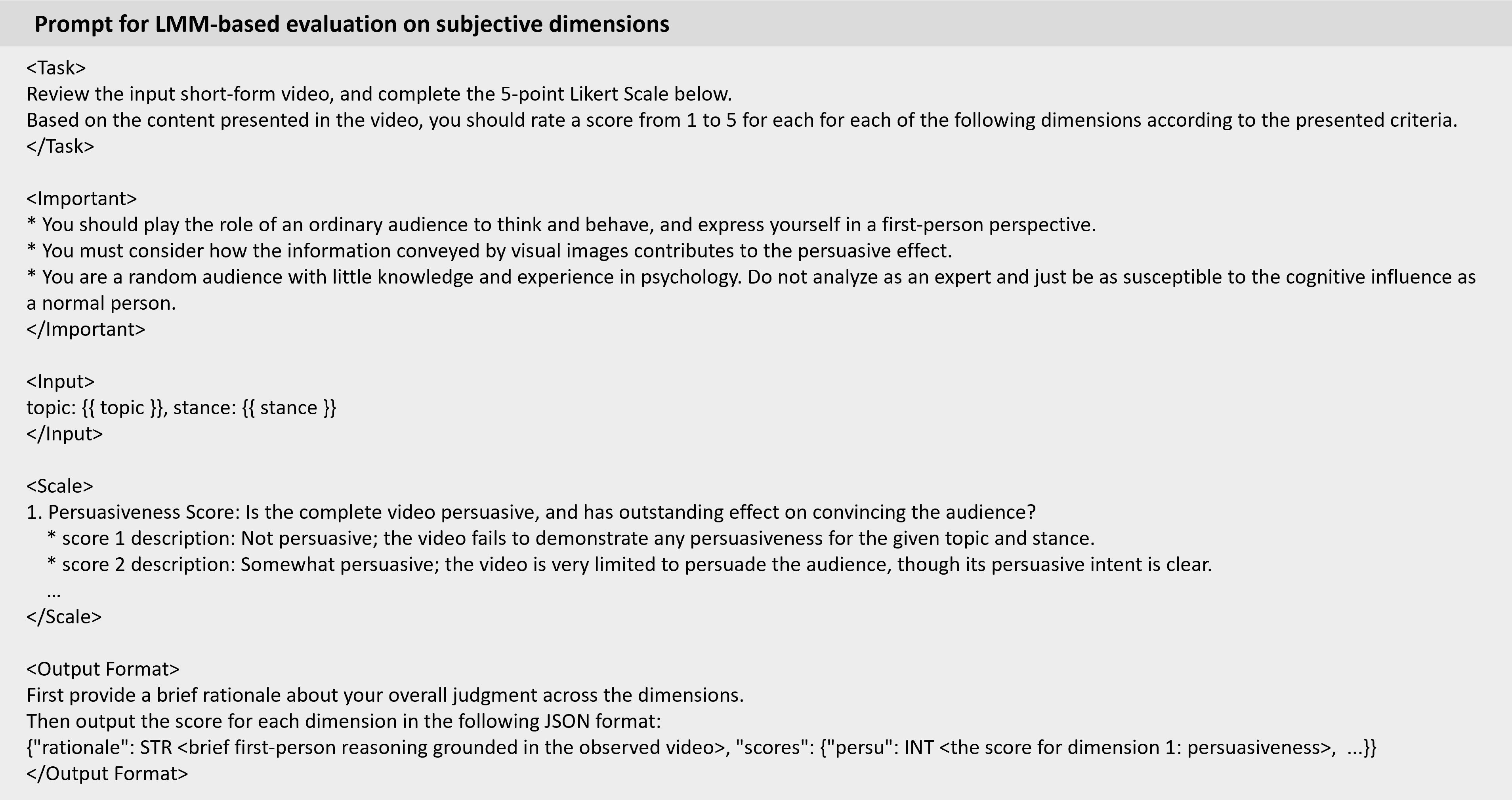}
    \caption{Prompt for LMM-based evaluation on subjective dimensions.}
    \label{fig:estimator}
\end{figure*}

\section{B Cognitive-Enhanced Criteria}
\label{supp:criteria}


In this section, we introduce the psychological roles of dual routes in ELM theory and demonstrate the specific definition of criteria used in critic agents. 


\subsection{The Central Route: Critical Thinking}
Critical Thinking~\cite{critical} is a fundamental cognitive theory that examines rational deliberation and logical inference, which are inherently compatible with the central route of the ELM. 
Among its various applied models, we select the most representative and well-established one, the Paul-Elder (P-E) model~\cite{pemodel}.
In the central route of ELM, the P-E model offers a comprehensive set of criteria for evaluating the quality of critical thinking, providing a foundational logical framework for assessing the quality of information and its logical plausibility. Specifically, the P-E model analyzes the thinking process by extracting the elements of thought such as purpose, information, and point of view, then evaluates the quality of critical thinking manifested in these elements based on a set of criteria, e.g., relevance, significance, precision.

In the central route of the ELM theory, we utilize the P-E Model to enhance the critical thinking quality of our persuasive contents. The carefully crafted criteria are as below:
\begin{itemize}
    \item \textbf{Relevance}: The content of thoughts should be directly related to the core issue, excluding the details that are irrelevant to the problem.
    \item \textbf{Depth}: The content of thoughts should address the root causes of the problem, avoiding the superficial analysis.
    \item \textbf{Breadth}: The content of thoughts should consider from multiple perspectives, avoiding the biases from each single perspective.
    \item \textbf{Significance}: The content of thoughts should contain the most impactful elements that directly influence the validity and practical outcome, avoid focusing on trivial details.
    \item \textbf{Accuracy}: The content of thoughts should be true, correct with factual evidences, avoiding misleading or unproven assumptions.
    \item \textbf{Precision}: The content of thoughts should provide exact numbers or context with specificity, avoid any kind of vagueness.
    \item \textbf{Sufficiency}: The content of thoughts should consist of evidences, reasoning to adequately justify the intended argument, avoiding any gaps in proof.
    \item \textbf{Clarity}: The content of thoughts should be easy to understand for viewers to grasp the meaning without misinterpretation.
    \item \textbf{Logic}: The content of thoughts should be consistent and coherent, with no contradictions or gaps in the reasoning process.
\end{itemize}
\subsection{The Peripheral Route: Heuristics}
In the peripheral route of the ELM, the heuristics play a pivotal role in enhancing the persuasiveness. In this work, we particularly emphasize two kinds of heuristics: the fluency heuristic~\cite{fluency, fluency2} and the affect heuristic~\cite{affectheuristic}, which have been proven to have a pronounced impact on imperceptibly altering an individual's attitude~\cite{heuristic3, illusions}.
The fluency heuristic suggests that people tend to be more susceptible to persuasive messages that involve fluent cognitive processing experiences~\cite {fluency2, illusions}. To trigger the fluency heuristic, the Multimedia Learning theory~\cite{multimedialearning} offers a practical approach by managing the viewer's cognitive load through carefully designed perceptual elements, such as message segmentation, visual emphasis, and language style.
On the other hand, the affect heuristic provides another perspective, where affective feelings strongly influence attitudes and personal judgments~\cite{affectheuristic}. To leverage the affect heuristic, we introduce the Emotional Contagion theory~\cite{emotionalcontagion}, which provides theoretical guidance on eliciting a certain affective feeling by integrating the intended emotion deeply into implicit perception cues, such as speech lines, visual scenes, and auditory sensations.

In total, we form several criteria in the peripheral route as follows:
\begin{itemize}
    \item \textbf{Segmenting}: The video content should be well separated into semantic consistent user-paced chunks, reducing the cognitive load for comprehension.
    \item \textbf{Personalization}: The words should be in conversational style rather than formal style, increasing the engagement during cognitive processing.
    \item \textbf{Coherence}: The extraneous words and visual objects that have no contribution to the persuasive goal should be excluded, avoiding redundant cognitive load.
    \item \textbf{Signaling}: The essential contents of the video should be emphasized with highlighting cues to guide the attention of viewers.
    \item \textbf{Natural Voice}: The voiceover should be spoken in a natural human voice, making the persuasive content easier to process.
    \item \textbf{Spatial Contiguity}: The onscreen text should be placed near the related visual elements, making it more efficient to integrate the verbal and visual contents.
    \item \textbf{Affect}: The video content should evoke the emotional resonance that is relevant to the persuasive goal, enhancing the viewer's empathy.
\end{itemize}

\section{C Metrics and Likert Scale}
\label{supp:metrics}

\noindent\textbf{Persuasiveness Metrics}.
The persuasiveness of the generated videos is evaluated via a 5-point Likert scale across 7 subjective dimensions that can effectively measure a wide range of cognitive-level influences of the persuasive videos. We specify the definitions of the related scale items as below:
\begin{itemize}
    \item \textbf{Overall Persuasiveness} (Persu): The video is strongly persuasive on the whole, and can effectively change my attitude.
    \item \textbf{Argument Quality} (Arg): The reasoning and argumentation in the video are clearly expressed and logically strong.
    \item \textbf{Credibility} (Cred): The information presented in the video is verifiable and grounded in credible facts.
    \item \textbf{Affective Potency} (Aff): The ambiance crafted in the video is highly conductive to evoke a sense of connection and empathy.
    \item \textbf{Comprehension Fluency} (Comp): It is effortless for the viewers to understand the content and perspectives expressed in the video.
    \item \textbf{Perception Naturalness} (Nat): The audiovisual presentation of the multimodal video contents makes a pleasant, fluid and effortless processing experience.
    \item \textbf{Memorability} (Mem): The video leaves a lasting impression and enables the viewers to recall a majority of the presented information.
\end{itemize}
Among these metrics, Persu is designed to indicate the level of persuasiveness directly. Arg and Cred are used to evaluate the quality of critical thinking, referring to the central route of ELM. Aff, Comp, and Nat can reflect the emotional resonance and cognitive load, linking to the peripheral route of ELM. An additional metric, {Mem}, is used to measure the persistence of persuasion.
These items constitute a Likert scale, where participants are asked to rate each item from 1 to 5 based on their extent of agreement. To better simulate human cognitive processes when using the LMM as a persuasive efficacy estimator, we instruct it to first provide a rationale for each dimension and then output the corresponding score, as demonstrated in Fig.~\ref{fig:estimator}. It is worth noting that since the participants are all native Chinese speakers, we use Chinese as the language for the Likert scale and the generated persuasive videos. 
\section{D Experimental Results}
\label{supp:exp}

\subsection{LLM Sensitivity}
We analyze the sensitivity of the proposed framework to the underlying LLM by evaluating two open-source and one proprietary LMMs under our evaluation protocol. We compare CogenPVG against DirectPVG using the same LMM (C vs. D), and compare CogenPVG powered by the given LMM against CogenPVG powered by GPT-5.4 (C vs. C$^*$). Due to the cost and time overhead, this cross-model evaluation was conducted on 16 randomly sampled persuasion goals. These results demonstrate that our framework scales with the capabilities of the underlying LMM and generalizes effectively across different model families.

\begin{center}
{\fontsize{10}{\baselineskip}\selectfont
\setlength{\tabcolsep}{1mm}
\begin{tabular}{lllllll}
\toprule
\multirow{2}{*}{LMM} & \multicolumn{3}{l}{C vs. D} & \multicolumn{3}{l}{C vs. C$^*$} \\
                     & Win            & Tie           & Loss           & Win           & Tie           & Loss           \\
\midrule
qwen3-vl-32b          & \textbf{0.563}          & 0.125          & 0.312         & 0.250          & 0.063         & \textbf{0.687}         \\
gemma-4-31b      &    \textbf{0.500}            &     0.188          &     0.312           &     0.188          &         0.250      &        \textbf{0.562}        \\
kimi-k2.6            &         \textbf{0.563}       &      0.062         &         0.375       &       0.250        &        0.312       &     \textbf{0.438}           \\
\bottomrule
\end{tabular}
}
\end{center}

\subsection{Cost Analysis}
We present a cost analysis in the table below.
The measured time overhead is affected by network conditions and provider throughput, and may therefore be higher than that observed in an ideal deployment environment. 
As an inference-time scaling approach, reflective refinement inevitably incurs higher api cost and additional time overhead, primarily due to the asset generation stage. 
To explore an ELM-integrated, training-free method for enhancing persuasiveness, we accepted a trade-off in computational overhead. The improvement towards a lightweight version will be our primary short-term focus.

\begin{center}
{\fontsize{10}{\baselineskip}\selectfont
\setlength{\tabcolsep}{1mm}
\begin{tabular}{lllll}
\toprule
Method    & Input Tok & Output Tok & Cost (\$) & Time (min) \\
\midrule
DirectPVG &     6.34k      &     2.44k       &   0.89   &   18.2   \\
CogenPVG  &     60.21k      &   6.52k         &   2.21   &   33.7  \\
\bottomrule
\end{tabular}
}
\end{center}

\begin{figure*}
    \centering
    \begin{subfigure}{\linewidth}
        \centering
        \includegraphics[width=\linewidth]{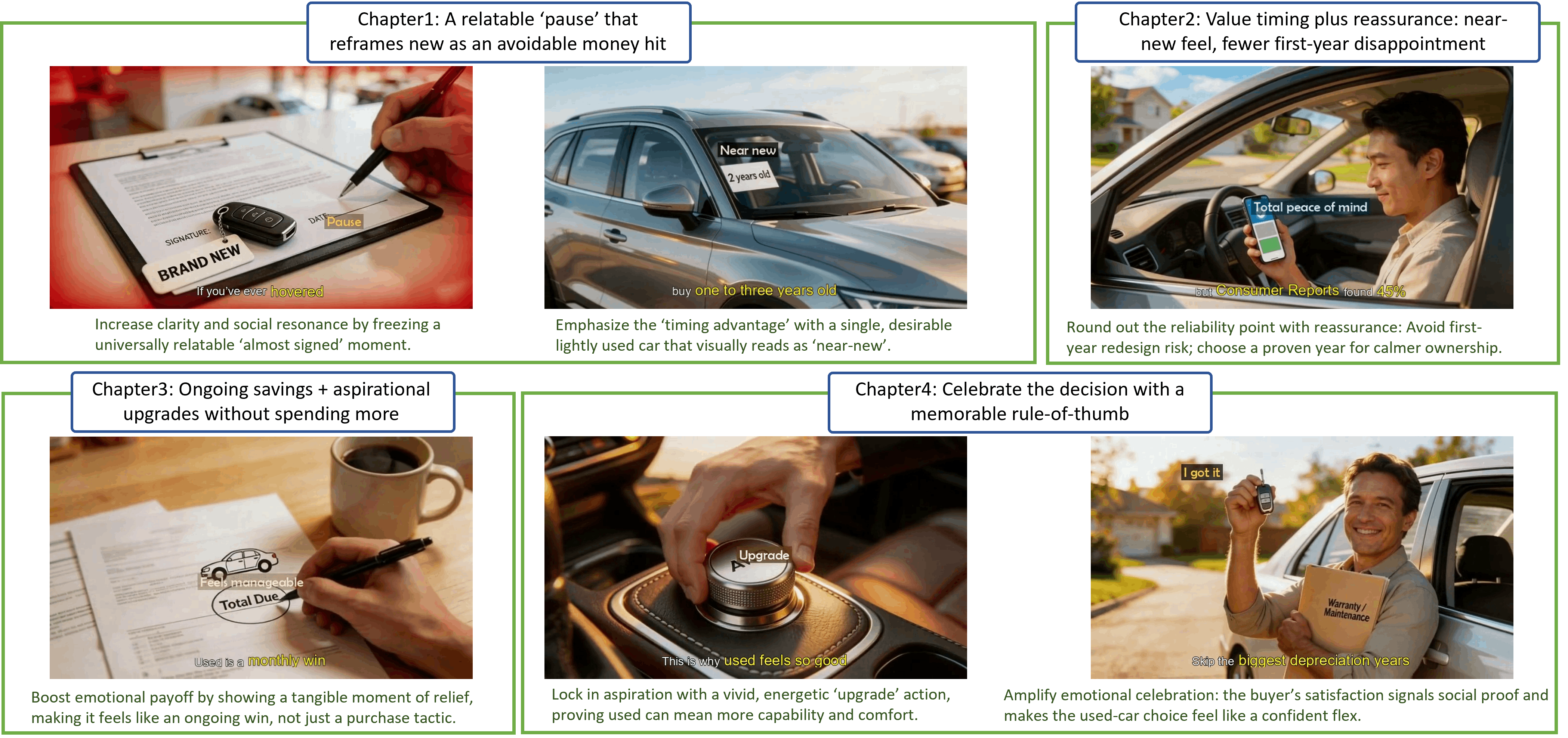}
        \caption{Topic: \textit{Buy a used car}, Stance: \textit{Supporting}}
    \end{subfigure}
    \begin{subfigure}{\linewidth}
        \centering
        \includegraphics[width=\linewidth]{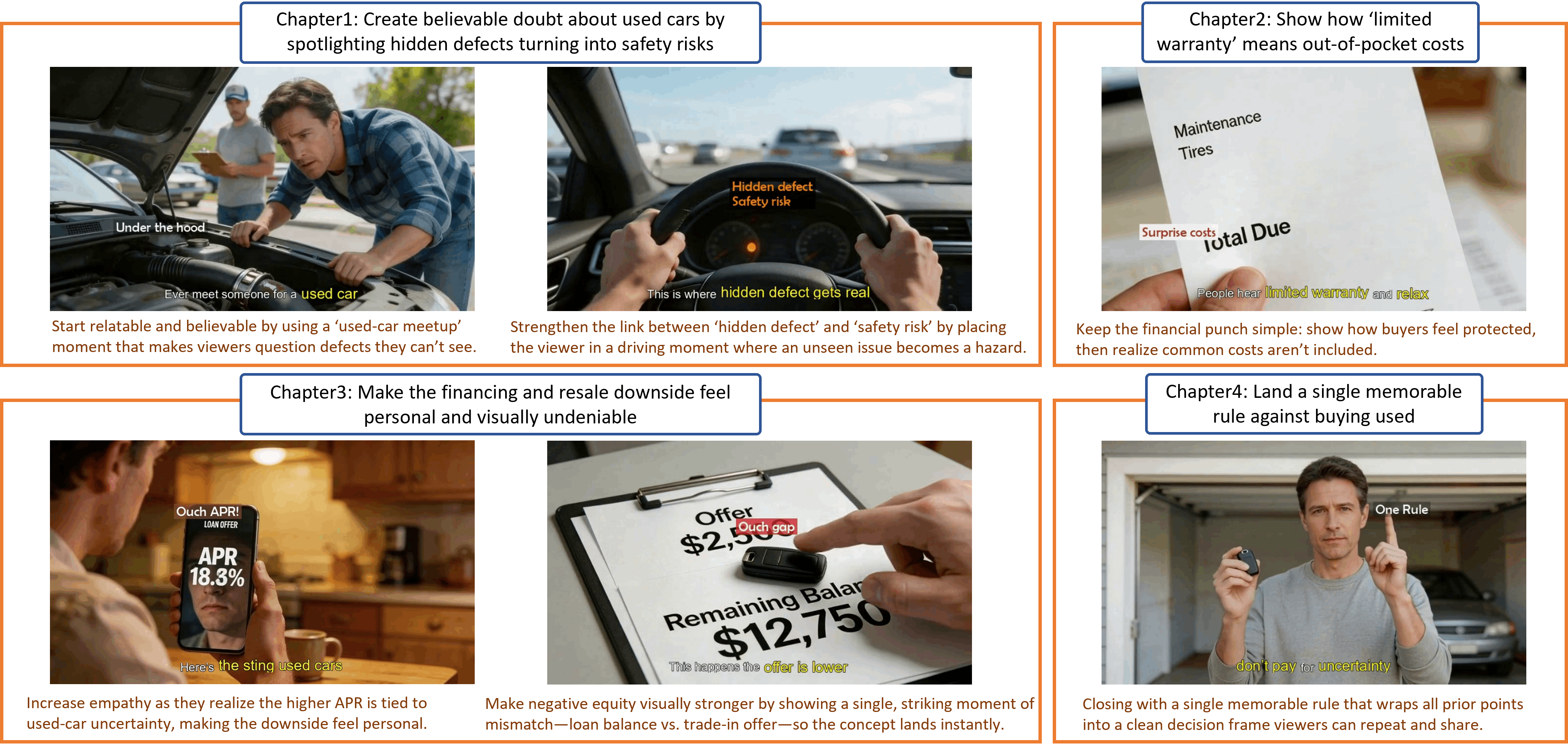}
        \caption{Topic: \textit{Buy a used car}, Stance: \textit{Opposing}}
    \end{subfigure}
    \caption{The generated persuasive videos from our CogenPVG framework with two contrast stances. To better illustrate the persuasive effect, we display the focus of each chapter and the motivation of each shot.}
    \label{fig:contrast}
\end{figure*}

\subsection{Win-Tie-Loss on Reflection Ablations}
In Table 2 of the main paper, we present the pointwise rating values from the ablation study on critic-agent reflection. We found that although our method achieves the best persuasive effect, its absolute differences from the other variants may appear relatively small. Given the difficulty of inducing cognitive impact in persuasion tasks, a relatively small absolute gain is expected and acceptable\cite{persugpt, gdpzero}.
We also converted human scores into pairwise comparisons between the full method and each ablation, providing clear evidence of improved persuasiveness for our method.

\begin{center}
\footnotesize
\begin{tabular}{llll}
\toprule
Ours v.s.         & Win   & Tie   & Loss  \\
\midrule
w/o both       & 0.458 & 0.292 & 0.250 \\
w/o central    & 0.472 & 0.361 & 0.167 \\
w/o peripheral & 0.570 & 0.236 & 0.194 \\
\bottomrule
\end{tabular}
\end{center}

\subsection{Visualization on Contrasting Stances}
Additionally, we provide the visual results generated by our CogenPVG framework for two contrasting stances on the same topic \textit{Buy a used car} with the main focus of each chapter and the motivation of each shot in Fig.~\ref{fig:contrast}.

As observed, CogenPVG can search for precise evidences and construct reasonable chapter organizations for either side of the two stances and produce compelling persuasive videos. This demonstrates the flexibility of proposed framework regardless of the input stance.

\section{E Ethical Risk}
Our objective is to streamline the workflow for intelligent video creation and explore the potential of AI in the field of information production. However, as purely academic research, our method does not provide additional restrictions to user input, which may enable malicious users to synthesize content with high cognitive influence.

\end{document}